\documentclass[a4paper,fleqn,usenatbib]{mnras}

\usepackage{newtxtext,newtxmath}
\usepackage{subfig}
\usepackage[T1]{fontenc}
\usepackage{ae,aecompl}

\usepackage{graphicx}	
\usepackage{amsmath}	
\usepackage{amssymb}	

\def\ut#1{\mathop{\vtop{\ialign{##\crcr
     $\hfil\displaystyle{#1}\hfil$\crcr\noalign
     {\kern1pt\nointerlineskip}\hbox{$\hfil\sim\hfil$}\crcr
     \noalign{\kern1pt}}}}}

\def\undersymbol#1#2{\mathop{\vtop{\ialign{##\crcr
     $\hfil\displaystyle{#2}\hfil$\crcr\noalign
     {\kern1pt\nointerlineskip}\hbox{$\hfil#1\hfil$}\crcr
     \noalign{\kern1pt}}}}}

\title[A Quasar microlensing event towards J1249+3449?]{A Quasar microlensing event towards J1249+3449?}

\author[F. De Paolis, A.A. Nucita, F. Strafella, D. Licchelli, G. Ingrosso]{
F. De Paolis $^{1,2}$\thanks{E-mail: depaolis@le.infn.it}
A.A. Nucita  $^{1,2}$\thanks{E-mail: nucita@le.infn.it}
F. Strafella $^{1,2}$\thanks{E-mail: strafella@le.infn.it}
D. Licchelli $^{3,4}$\thanks{E-mail:domenico.licchelli@le.infn.it}
G. Ingrosso $^{1,2}$\thanks{E-mail: ingrosso@le.infn.it}
\\
$^{1}$ Department of Mathematics and Physics {\it ``E. De Giorgi''} , University of Salento, Via per Arnesano, CP-I93, I-73100, Lecce, Italy\\
$^{2}$  INFN, Sezione di Lecce, Via per Arnesano, CP-193, I-73100, Lecce, Italy\\
$^{3}$ R.P. Feynman Observatory, I-73034, Gagliano del Capo, Lecce, Italy  \\ 
$^{4}$ CBA, Center for Backyard Astrophysics - I-73034, Gagliano del Capo, Lecce, Italy\\
}

\date{Accepted XXX. Received YYY; in original form ZZZ}

\pubyear{2017}

\begin{document}
\label{firstpage}
\pagerange{\pageref{firstpage}--\pageref{lastpage}}
\maketitle

\begin{abstract}
We show that the optical flare event discovered by \cite{Graham2020} towards  the active galactic nucleus J1249+3449 is fully consistent with being a quasar microlensing event due to a $\simeq 0.1 M_{\odot}$ star, although other explanations, such as that, mainly supported by \cite{Graham2020}, of being the electromagnetic counterpart associated to a binary black hole merger, cannot be completely excluded at present.
\end{abstract}




\section{Introduction}
Very recently  \cite{Graham2020}  reported the detection of an electromagnetic flare event, a bump in the optical light-curve of the active galactic nucleus (AGN)  J1249+3449, which has been observed for 21.4 months by the Caltech's robotic Zwicky Transient Facility (ZTF), a dedicated survey telescope which employs a 47 deg$^2$ field-of-view camera on the Palomar 48-inch Samuel Oschin Schmidt Telescope.

Very interestingly, about 47 days before the bump peak (see also the  next section), at 2019-05-21, 03:02:29 UTC,  a gravitational wave (GW) event, named S190521g, was detected by both the LIGO detectors and the VIRGO detector, with an estimated false alarm rate (FAR) $\simeq 3.8\times 10^{-9}$ Hz (corresponding to a rate of one false event every about 8.3 yrs)  \citep{LIGO2019}. The estimated luminosity distance was of $3931 \pm 953$ Mpc and the event was classified as a binary black hole (BBH) merger with a  probability of about $97\%$.

The flare event, beginning at about MJD=58650, lasted for about 40 days, and is characterized by a $\simeq 5\sigma$ departure with respect to the ZTF baseline for this source.
The authors of this important discovery have considered many possibilities to account for this peculiar observation: intrinsic variability of the AGN, a supernova or a tidal disruption event, a microlensing event and an electromagnetic counterpart associated to the binary black hole merger, the last one being, maybe, the most challenging option. In this scenario, when the two black holes merged, the new one experienced a {\it kick} that sent it flying off in a random direction within the gaseous accretion disk of the supermassive black hole, giving rise to the bright electromagnetic flare  (see, e.g., \citealt{McKernan}).

In any case,  \cite{Graham2020}  dismissed all other hypotheses, but the last one, on the basis of their negligible estimated probabilities, of order of $O(10^{-3})$ or less. 

The aim of the present Letter  is to show that the flare event in question is  completely consistent with a microlensing event,  maybe it is the most clear and definite of such events towards a quasar to date, and the obtained parameters, as well as the optical depth and the microlensing rate values,  are consistent with the expectations for such kind of events.

\section{Data analysis and results}
As mentioned in the previous Section, the flare event observed in J1249+3449 closely 
resembles a Paczyński bump which characterizes a typical microlensing event. Of course, 
in order to advance the hypothesis of a microlensing nature, the event must fulfill 
three minimal requirements: {\it a)} the observed light curve should be achromatic since the gravitational bending of light is independent of the frequency of the radiation; {\it b)} for a single lens event, the observed bump should be symmetric with respect to its maximum and the overall shape must be dependent on only four parameters, namely the Einstein crossing time $t_E$, the impact parameter $u_0$, the time of closest approach $t_0$ and the magnitude of the source baseline (not amplified), i.e. far in time with respect to the microlensing event; and, {\it c)} any source experiencing a microlensing event should be characterized by a constant flux outside of the event, and should not have any periodic feature with the same strength as the  
event at maximum and/or similar time-scales.

\begin{figure*}
  \centering
 \includegraphics[width=0.42\textwidth]{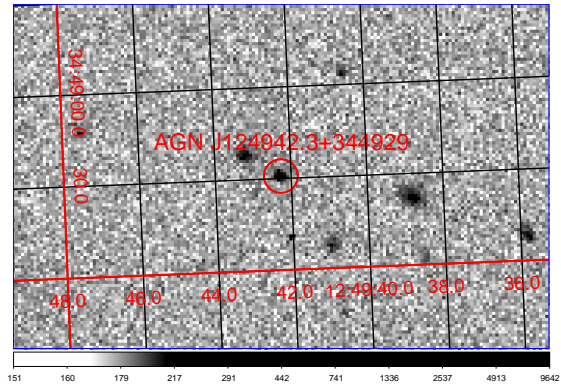}
\caption{The Agn J1249+3449 observed in the $r$ band by using the ZTF instruments.}
\label{fig_ds9}
\end{figure*}

\begin{figure*}
  \centering
  \subfloat[]{\includegraphics[width=0.62\textwidth]{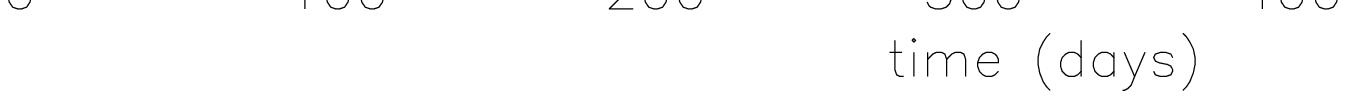}\label{fig2a}}
  \hfill
  \subfloat[]{\includegraphics[width=0.62\textwidth]{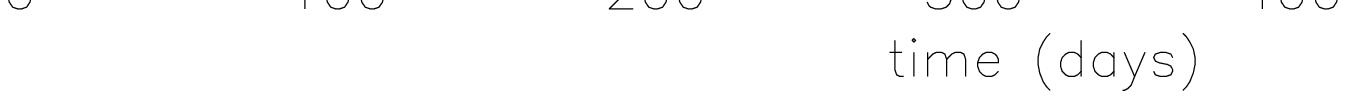}\label{fig2b}}
  \caption{In panel (a) we give the $g$ and $r$ band ZTF data with green and red dots, respectively. Here  the data are scaled as described in the text and the black line represents the best fit microlensing model. Here, we scaled the time axis to the start of observation. The bottom panel shows the model residuals, clearly indicating the absence of any trend.}
  \label{bestfit}
 \end{figure*}

Hence, in order to test the microlensing hypothesis we retrieved the available data (only $r$ and $g$ band light curves, being the $i$ band characterized by too sparse data points) from the ZTF database (\url{https://www.ztf.caltech.edu}). In Figure \ref{fig_ds9}, we give a zoom around the source detected in the ZTF $r$ band.  
{After screening the data sets from measurements flagged as saturated, we were left with data covering the observing time (in modified Julian day, MJD) from $58202.293102$ (re-scaled as $t=0$ in Fig. \ref{bestfit}) to $58846.551227$ corresponding to calendar dates 2018-03-25, 07:02:04.013 UTC and 2019-12-29, 13:13:46.013 UTC, respectively. The available light curve then lasts for $\simeq 1.8$ years with the peak of the event observed at $\simeq$ MJD 58671.628 (2019-07-07 15:04:50.995 UTC). Hence, the bump peak occurred  $\simeq 47$ days after the LIGO/VIRGO interferometers detected the gravitational wave event (which occurred at 2019-05-21 03:02:29 UTC).}

We first noted that the  $g-r$ color remained constant, within the uncertainties, during the flare event.  This is an hint that the bump is achromatic, as requested by the microlensing event assumption, so that we fix the color to the differences between the median values outside the event, i.e. $\simeq 0.17$. Hence, in order to account later on for all the information contained in the data, we scaled the $r$ band data to the $g$ data (as usually done in microlensing studies) {by considering the constancy of the $g-r$ color over the event duration}. In the following, we scaled the time axis to zero in order to have more manageable numbers.

Since the bump resembles a microlensing event, we quantitatively checked this hypothesis  by fitting a Paczyński (single lens) light curve to the data. In particular, the expected flux amplification $A(t)$ in this case is 
\begin{equation}
    A(t)=\frac{u^2(t)+2}{\sqrt{u^2(t)(u^2(t)+4)}}
\end{equation}
where $u^2(t)=(t-t_0)^2/t_E^2 +u_0^2$ is the projected distance in units of the Einstein radius between the lens and source {(for details see, e.g., \citealt{Schneider})}. Switching to logarithm and adding a constant magnitude $m_b$ to account for the source baseline, the final adopted model results to be
\begin{equation}
  m(t)=-2.5\log[A(t)]+m_b
\end{equation}
The model was fitted to the data leaving all the parameters free to vary and starting the Levenberg-Marquardt minimization procedure from an adequate set of initial parameter values (see, e.g., \cite{Lee2009} and references therein). In panel (a) of Figure \ref{bestfit} we give the best-fit light curve (black line) over-plotted to the ZTF  data points of J1249+3449 with, in particular, the red and green dots corresponding to the $r$ and $g$ bands (opportunely scaled as described above), respectively. In panel (b) of the same Figure we present the model residuals which show an average value consistent with zero, without any trend or departure from being flat.
 The best fit parameters  (we note that the best fit is remarkably good, with $\chi^2\simeq 1.2$ for 509 d.o.f. and  1$\sigma$ confidence level errors) are $t_0=469\pm 1$ days, $t_E=17.6\pm 1.2$ days, $u_0=0.92\pm0.03$ and $m_b=19.082\pm0.003$, respectively. The corresponding amplification at maximum of the event is $\simeq 1.41$, i.e. quite similar to what expected for a randomly-chosen microlensing event. Note also that, by considering the typical scatter of the baseline data, the event peak corresponds to a $\simeq 5 \sigma$ signal. Therefore,  the probability that this is due to a random fluctuation of the source is negligible (it is certainly less than 1 over $3.5\times 10^6$). One can note the presence of three {\it isolated} data points at $\simeq 520$ days from the observation start, on the descending part of the  event light curve. Although one could think to an asymmetry in the light curve, these data are in a poorly sampled region and well within the typical scatter of this source which, by the way, is also characterized by a flickering on the time scale of a few days.  Furthermore, also the fit in Fig. 2 of \citet{Graham2020} does not return a better match to these data. Only a better sampling of the descending part of the light curve would have allowed a compelling evaluation of this issue.

The GW observation data showed that the luminosity distance of the J1249+3449 source is $3931 \pm 953$ Mpc (\citealt{LIGO2019}). Therefore, if the bump was due to a microlensing event, the intervening lens must be located in the same host galaxy. As usual, the Einstein time of the event is
\begin{equation}
t_E=\frac{2R_E}{v_{\perp}}=2\frac{ \sqrt{\frac{4Gm}{c^2} \frac{D_LD_{LS}}{D_S} }}{v_{\perp}}
\end{equation}
where $R_E$ is the linear Einstein radius, $m$ is the lens mass, $v_{\perp}$ the component of its velocity orthogonal to the line of sight, and $D_L$, $D_S$ and $D_{LS}$ are the lens, source and relative source-lens distances, respectively.
With the approximation of $D_S\simeq D_L$ and solving with respect to $m$ one has
\begin{equation}
m=\frac{v^2_{\perp} t_E^2 c^2}{16 G D_{LS}}
\end{equation}
Assuming for the AGN host galaxy stars a typical velocity in the range 200-400 km s$^{-1}$ (see next) and a source-lens distance $D_{LS}$ in the range $1-10$ kpc, the lens mass value results to be  in the range $0.01-0.5$ M$_{\odot}$, again consistent with the typical values of the expected lens mass.  

Let us next evaluate the optical depth and the rate for such microlensig event under reasonable assumptions. As described in \citet{dunlop2003}, who studied 
a sample of radio-quiet (RQQs) and radio-loud (RLQs) quasars as well as radio galaxies (RGs) observed in the Hubble Space Telescope (HST) deep images, a quasar host galaxy can be well described by a $r^{-\beta}$ profile with exponent in the range $0.19-0.26$, with $\beta=0.25$ corresponding to the classical de Vaucouleurs profile. Furthermore, the typical scale-length of the projected luminosity is $R_e\simeq 10$ kpc. We then use this information to model 3D galaxy hosting the AGN J1249+3449. However, since the de Vaucouleurs profile cannot be de-projected analytically, we used the Hernquist approximation \citep{hernquist1990} for the observed surface brightness profile. The advantage is that the profile can be de-projected in a 3D mass-density profile given by
\begin{equation}
    \rho(r)=\frac{M}{2\pi} \frac{a}{r}\frac{1}{(r+a)^3}
\end{equation}
corresponding to the mass  profile 
\begin{equation}
    M(r)=M \frac{r^2}{(r+a)^2}
\end{equation}
where $M$ is the total galaxy mass in the form of stars, and $a\simeq R_e/1.8153$ is a scale length parameter (here $R_e$ is the effective radius corresponding to the isophote enclosing half of 
the galaxy's luminosity). As usual,  we consider a cut-off radius $R$ in the $\rho (r)$  profile, with $R\simeq 10 a$.

With the adopted model, the optical depth for microlensing event reads out 
\begin{equation}
\tau=\int_0^R \pi R^2_E(r) \frac{\rho(r)}{m}dr=\frac{2GM}{c^2}a\left[-\frac{1}{2(R+a)^2}+\frac{1}{2a^2}\right]   
\label{tau}
\end{equation}
where the integration is performed over any possible distance of the lens in the host galaxy. By fitting the $H\beta$ line profile of the AGN, \cite{Graham2020} estimated a super massive black hole of mass in the range $(1-10)\times 10^8$ M$_{\odot}$ which corresponds, for a standard black hole mass/bulge mass relation (see, e.g. \citealt{Merritt2001,Jahnke2009}) to a bulge mass of $M_{\rm bulge}\simeq  10^{11}-10^{12}$ M$_{\odot}$.  
The obtained microlensing  optical depth for AGN J1249+3449 from eq. (\ref{tau}) turns out to be $\tau\simeq  10^{-6}-10^{-5}$, which can be considered as a lower limit to the real value of $\tau$, as we considered only the bulge component of the galaxy.

The microlensing rate, i.e. the number of events expected to be detected per quasar and per year, can be evaluated as
\begin{equation}
\Gamma=\int_0^R 2 R_E (r) \frac{\rho(r)}{m} \sqrt{2} \sigma(r) dr
\label{gamma}
\end{equation}
where the expression for the 1-dimensional dispersion velocity $\sigma(r)$ is given in  \citet{hernquist1990}. Evaluating numerically the previous integral we get a microlensing rate $\Gamma\simeq 2\times 10^{-4} (m/0.1 M_{\odot})^{-0.5}$ events quasar$^{-1}$ yr$^{-1}$.
We note that for the adopted model, the rate scales with the mass of the host galaxy as $M^{3/2}$.

We would like to stress here that the value of $\Gamma$ estimated above based on the Hernquist model  should be regarded as a conservative lower limit. Indeed, the microlensing rate  clearly depends on both the total mass of the host galaxy component able to act as microlenses and on their transverse velocity. By using the fundamental plane relations for the  supermassive black hole mass/Bulge mass/stellar dispersion velocity and assuming a supermassive black hole mass of $10^8-10^9$ M$_{\odot}$ one finds M$_{\rm bulge}=10^{11}-10^{12}$ M$_{\odot}$ and $1-$dimensional dispersion velocity with lower bound of about 200 km s$^{-1}$ and upper bound in the range $300-340$ km s$^{-1}$ (see, e.g. \cite{Ferrarese2000,Gebhardt2000,Merritt2001,Jahnke2009}). With these values the microlensing rate turns out to lie in the range $6\times10^{-5}$--$1\times 10^{-3}$ events quasar$^{-1}$ yr$^{-1}$, for the adopted lens mass of 0.1 M$_{\odot}$.
 We would like to notice here that our estimate of the microlensing event rate is not in contradiction  with that in \cite{Graham2020} ($\Gamma\simeq 10^{-4}$) where a population of about $10^{10}$ stars (with mass $\simeq 1$ M$_{\odot}$) in the host galaxy was assumed. Note also that the event rate scales with $m^{-1/2}$, M$^{3/2}$ and is also proportional to the stellar dispersion velocity $\sigma$.

\section{Conclusions}
We have found that the optical flare event detected about 47 days after the GW event S190521g towards the AGN J1249+3449 appears to be fully consistent with a quasar microlensing event, while the authors of discovery of the event attributed it to the more challenging possibility that it is the first electromagnetic counterpart to the binary black hole merger, immersed in the supermassive black hole accretion disk. Indeed, we find that the available ZTF data in the $g$ and $r$ bands are perfectly in agreement with a microlensing Paczyński light curve, and show also a remarkable achromatic behavior, as expected in the case of a microlensing event. Also the other obtained microlensing parameters, $t_E=17.6\pm 1.2$ days, $u_0=0.92\pm0.03$ and the peak magnification $A_{\rm max}\simeq 1.41$, appear to be typical parameters for a microlensing event. Moreover, the microlensing rate has been estimated to be in the range  $\Gamma\simeq 6\times 10^{-5}-1\times 10^{-3}$ events quasar$^{-1}$ yr$^{-1}$, implying that the probability of occurrence \footnote{The rate $\Gamma$ gives directly the probability of occurrence of a microlensing event per year towards the considered quasar.} of such events towards AGN J1249+3449 is absolutely not negligible, also taking into account that the quasar in question has been monitored for more than  15 years.

Even if  other explanations cannot be  ruled out at present, we would like to emphasize that the observation, within a few years, of an optical flare similar to that analyzed here, would strongly support the \cite{Graham2020} model \footnote{We note that \cite{Graham2020} predicted within their scenario a flare repetition rate $\simeq 1.6 {M_8}a_3^{3/2}$ yr, where $M_8$ is the supermassive black hole mass in units of $10^8 M_{\odot}$ and $a_3$ is the accretion disk radius in units of $10^3$ times the supermassive black hole gravitational radius.}, while in the case of the microlensing nature of the event no repetition is expected. 

We would also like to observe that our study, and in particular the estimated rate for microlensing events per quasar and per year, is relevant in the context of the next-coming large surveys of the sky, such as that provided by the Vera C. Rubin Observatory with the Legacy Survey of Space and Time (LSST) survey, which will soon produce a survey of the whole Southern sky every a few days for a decade \cite{LSST}. In this respect,  a very recent theoretical study by \cite{Neira2020} presented a tool to generate mock quasar microlensing light curves. It is noticeable the finding that the expected duration of the majority of the high-magnification events is between 10 and 100 days. This provides an additional evidence that the flare event observed towards the AGN J1249+3449  is consistent with  a microlensing event. 

Note also that LSST survey will allow to detect a sample of at least $10^7$ AGNs and quasars up to a limiting magnitude of $\simeq 24.1$ and for redshifts $z<2.1$ \citep{lsstsciencebook}. Therefore, with the above microlensing rate estimate   (even if we caution that it was derived for a particular AGN), the expected number of similar microlensing events can be as large as $6\times 10^2$--$10^4$. \footnote{With the same rate we expect between 36 and 300 microlensing events per year for the ZTF survey which detected $\simeq 6\times 10^5$ spectroscopically confirmed quasars \citep{Assef2018} and turns out to be in agreement with the 209 ZTF alerts at $5\sigma$ above the background and duration longer than 20 days (again we caution that the rate calculated towards a particular quasar has been used here as an average value). This issue certainly deserves a deeper investigation.} Furthermore, since the LSST is designed to observe the whole Southern sky every a few nights,  the survey could trigger prompt observations (possibly with larger cadence) of interesting targets.

Detecting such kinds of transient events and characterizing them to infer their nature is a very challenging field of research and requires a very high cadence, as well as good photometric stability, to capture and resolve them. This may  certainly be obtained by LSST and/or other facilities which will be available in the near future.

\section*{Acknowledgements}
 We thank for partial support the
INFN projects TAsP and EUCLID.

\section{Data availability}

The observational data underlying this article were accessed from the Zwicky Transient Facility (ZTF) \url{https://www.ztf.caltech.edu}. The derived data generated in this research will be shared on reasonable request to the corresponding author.





\bsp	
\label{lastpage}

\begin{thebibliography}{99}
\bibitem[\protect \citeauthoryear{LSST Science Collaboration}{2009}]{lsstsciencebook} 
Abell,P.A. et al., 2009,  arXiv e-prints: 0912.0201v1 

\bibitem[\protect \citeauthoryear{Assef et al.}{2018}]{Assef2018}
Assef, R.J, Stern, D., Noirot, G. et al., 2018, ApJS 234, 23

\bibitem[\protect \citeauthoryear{Graham et al.}{2020}]{Graham2020}
Graham, M.J. et al., 2020, Phys. Rev. Lett. 124, 251102,  (preprint arXiv:2006.14122v1)

\bibitem[\protect\citeauthoryear{LIGO/VIRGO}{2019}]{LIGO2019}LIGO/VIRGO Collaboration, 2019, GCN Circular 24621

\bibitem[\protect \citeauthoryear{Dunlop et al.}{2003}]{dunlop2003}
Dunlop, J.S., et al., 2003, MNRAS, 340, 1095

\bibitem[\protect \citeauthoryear{Ferrarese \& Merritt}{2000}]{Ferrarese2000} Ferrarese, L., Merritt, D., 2000, ApJ, 539, L9

\bibitem[\protect \citeauthoryear{Gebhardt et al.}{2000}]{Gebhardt2000} Gebhardt, K., Bender, R., Bower, G., et al., 2000, ApJ, 539, L13

\bibitem[\protect \citeauthoryear{Hernquist}{1990}]{hernquist1990}
Hernquist, L., 1990, ApJ, 356, 359

\bibitem[\protect \citeauthoryear{Jahnke et al.}{2009}]{Jahnke2009} Jahnke, R., Bongiorno, A., Brusa, M. et al., 2009, Apj, 706, L215

\bibitem[\protect \citeauthoryear{Lee et al.}{2009}]{Lee2009} Lee, C.-H., Riffeser, A, Seitz, S., Bender, R., 2009, ApJ, 695, 200

\bibitem[\protect \citeauthoryear{LSST}{2017}]{LSST} LSST Science Collaboration, Marshall, P., Anguita, T., et al. 2017, arXiv e-prints, arXiv:1708.04058

\bibitem[\protect \citeauthoryear{McKernan et al.}{2019}]{McKernan} McKernan, B.,  Ford, K. E. S., Bartos, I. et al., 2019, ApJL 884, L50

\bibitem[\protect \citeauthoryear{Merritt \& Ferrarese}{2001}]{Merritt2001}
Merritt, D. \&, Ferrarese, L.,  2001, MNRAS, 320, L30

\bibitem[\protect \citeauthoryear{Neira et al.}{2020}]{Neira2020} 
Neira, F., Anguita, T. and Vernardos, G., 2020, MNRAS, 495, 544

\bibitem[\protect \citeauthoryear{Schneider et al.}{1992}]{Schneider} Schneider, P., Ehlers, J., Falco, E. E.,  Gravitational Lenses, 1992, Springer-Verlag, Berlin

\end{thebibliography}
\end{document}